\documentclass[preprint,12pt]{elsarticle}

\usepackage{graphicx,epsfig}
\usepackage{amsmath}
\usepackage {amssymb}
\usepackage{color}
\usepackage[T1]{fontenc}
\usepackage[utf8]{inputenc}

\usepackage{lineno}
\journal{Journal Name}

\begin{document}

\begin{frontmatter}


\title{ Quantization of the interior of the black hole}


\author[label1]{Laysa G. Martins}
\ead{laysamartinsymail@yahoo.com.br}
\address[label1]{Departamento de Física, Universidade Federal de Juiz de Fora (UFJF), Juiz de Fora-MG, CEP 36036-900, Brazil.}
\author[label2]{K. Luz-Burgoa}
\ead{karenluz@ufla.br}
\address[label2]{Departamento de Física (DFI) and Museu de Historia Natural (MHN), Universidade Federal de Lavras (UFLA), Lavras-MG,Caixa postal 3037, CEP 37000-900, Brazil.}
\author[label2]{Jos\'{e} A. C. Nogales}
\ead{jnogales@ufla.br}

\begin{abstract}
In this work we study the Schwarzschild metric in the context of canonical quantum gravity inside the horizon, close of horizon and near the black hole singularity. Using this standard quantization procedure,  we show that the horizon is quantized and the black hole singularity disappears. For the first case, quantization of the Schwarzschild radius was obtained in terms of the Planck length $l_{Pl}$, a positive integer $n$ and the ordering factor of the operator $p$. From the quantization of the Schwarzschild radius it was possible to determine the area of the black hole event horizon, its mass and the quantum energy of the Hawking radiation as well as its frequency. For the solution close to the interior black hole singularity, the wave function was determined and applied the DeBroglie-Bohm interpretation. The Bohm's trajectory was found near to the singularity. It which describes how the spacetime evolves over time and depends on the ordering factor of the operator $p$. Thus, for the case where $|1-p|\neq0,3$, the Bohm's trajectory is finite and regular, that is, the singularity is removed. For the case where $|1-p|=3$, the Bohm's trajectory assumes an exponential behavior, never going to zero, avoiding the singularity.That result allows that spacetime be extended beyond the classical singularity.

\end{abstract}


\end{frontmatter}

%\linenumbers

\section{Introduction}
\label{S:1}
\indent The knowledge of the space-time structure, at a fundamental level, is an issue in development. One of the important predictions of the Einstein's theory of general relativity is the formation of space-time singularities and the nature of black hole horizons. There is a great conviction that this problem will be solved by a quantum theory of gravity.\\
\indent Meanwhile, spherically symmetric minisuperspaces are an interesting \hyphenation{a-re-na}arena to test and clarify ideas of quantum gravity. The are also, in order to avoid the technical problems resulting from the quantization of the complete gravitational field. In the minisuperspace models, we restrict the gravitational and matter fields to be homogeneous.\\
\indent Quantum singularities were studied for different  situations and \hyphenation{ge-ne-ra-li-za-tions}generalizations, in particular singularities inside of black holes \cite{konkowski2001quantum,helliwell2003quantum,konkowski2003quantum,doi:10.1142/S0217751X11054334,pitelli2007quantum,pitelli2008quantum,pitelli2009quantum,letelier2010n,unver2010quantum,mazharimousavi2009generating,mazharimousavi20112+,gurtug2012quantum}. Black holes properties can be described with Einstein’s theory of gravity however, there are several aspects that are currently not well understood, for that reason  it is expected that they will require an extension of the classical theory. In particular the interior of a black hole has been widely studied by several authors, from the classical and quantum point of view, however there are still many unsolved problems. For instance, there are many ambiguities in solutions such as, the presence of an event horizon, or a naked singularity or hidden behind the event horizon.\\
\indent In this work a non-probabilistic interpretation will be adopted for studying the singularity inside of horizon, so it will not be necessary to have a measuring apparatus or a classical domain to recover physical reality. In this sense, we will use DeBroglie-Bohm interpretation of quantum mechanics, which has been applied to many models \cite{de1998causal,de1998causal1,oliveira2018broglie}.\\
\indent  In this communication we study the interior of a Schwarzschild black hole, close to the singularity and the horizon. In particular we use the canonical quantization, Arnowitt-Deser-Misner (ADM) formalism, in a \hyphenation{par-ti-cu-lar}particular spherically symmetric vacuum space-times, in a appropriate coordinate system. We focus on the Schwarzschild solution inside and close to the horizon and near the singularity, considering the \hyphenation{or-de-ring}ordering factor in the moment operator. This allows us to obtain the Bohm's equation of quantum evolution, with which we can study the singularities completely, that is, we show that the black hole singularity disappears and the horizon is quantized, as done in some works \cite{makela1996schroedinger,modesto2004}.\\
\indent The attempts to quantize Einstein's theory of gravitation in the \hyphenation{ca-no-ni-cal}canonical framework was overshadowed by the existence of constraints among the dynamical operators. While the consistency of the classical constraints is ensured by the first-class algebra obeyed by them, the realization of the corresponding quantum algebra is obstructed by ambiguities in the ordering of dynamical operators. Without one guiding principle, the quantization of dynamical variables is subject to factor-ordering ambiguities \cite{kuchar1993canonical,brahma2015spherically,vilenkin1999factor,dewitt1967quantum,tsamis1987factor}. \\
\indent Some authors have come to the conclusion that there exists solution to the factor-ordering ambiguities and they proposed a large class of solutions where those ambiguities restrict other properties of the system. Unless one knows how to solve the factor-ordering ambiguities, one really does not know how to construct quantum theory from this point of view. In a sense, the right factor ordering in the quantum theory i.e. a consistent factor ordering of dynamical operators is a remarkable unsolved problem. In this article we will consider the ordering factor in the dynamic operators without any restriction. \\
\indent This work is organized as follows: in section 2 the metric of the \hyphenation{Schwarzs-child}Schwarzschild black hole inside the event horizon is presented. In section 3, we find the dynamic equations for the interior of the black holes, using the Arnowitt-Deser-Misner (ADM) formalism, which decomposes spacetime into space + time, so that the tensor fields are worked only in three-dimensional space (hypersurface). In section 4, the quantization inside the black holes is made, considering the ordering factor of the operators. In section 5, the solutions to the Schr\"odinger equation for black holes are presented, and we study the regions close to the event horizon and close to the interior singularity. For both cases, the singularities are shown to disappear. Finally, in section 6, some conclusions and comments are presented.

\section{The Black Hole metric inside the horizon}
Following \cite{modesto2004}, let's start by considering the Schwarzschild solution given by:
\begin{eqnarray}
ds^2 = - \frac{c^2 dT^2}{\Big(\frac{2 M G}{c^2 T } -1 \Big)} + \Bigg(\frac{2
M G}{c^2 T} -1 \Bigg)  dR^2 + T^2 d\Omega^2,
\end{eqnarray}
where $M$ is the mass of Schwarzschild black hole, $G$ corresponds to the gravitational constant, $c$ is the speed of light and $d\Omega^2=\sin^2 \theta d\phi^2 + d \theta^2$ corresponds to a solid angle element. For $T<2MG/c^2$ this metric describes the space-time inside the horizon of a Schwarzschild metric \cite{modesto2004}. The coordinate $T$ is timelike and the coordinate $R$ is spatial-like with $T \in ] 0 , 2 M G[$ and $R \in ] - \infty
, + \infty[$. We can define  a new temporal variable $\tau$ by $c^{2}$
\begin{equation}
d \tau = \frac{dT}{\sqrt{\frac{2 M G}{c^2 T} -1}}.
\label{tempo}
\end{equation}
The integration gives 
$
\tau = - \sqrt{T(2 M G/c^2 -T)} + (2 M G/c^2) \arctan \Bigg(\sqrt{\frac{T}{2
M G/c^2 -T}}\Bigg) + \textrm{const}$.
We take $\textrm{const} = 0$, because $\mbox{lim}_{T \rightarrow 0} \, \tau(T) =\textrm{const}$.  The function $T = T(\tau)$ is monotonic and convex, thus $\tau
\in ] 0 , (2 M G/c^2) \pi /2[$.  In this new temporal variable the metric becomes
\begin{eqnarray}
ds^2 = - c^2 d \tau^2+ \Bigg(\frac{2 M G}{c^2 T(\tau)} -1 \Bigg) dR^2 +
T(\tau)^2 (\sin^2 \theta d\phi^2 + d \theta^2).
\end{eqnarray}
\indent Then, we make a new change of variables introducing two function $b^2(\tau) \equiv \frac{2 M G}{c^2 T(\tau)} -
1$ and $a^2(\tau) \equiv T^2(\tau)$, also redefine $\tau \equiv t$ and $R \equiv r$.  The metric become
\begin{eqnarray}
ds^2 = - c^2 d t^2+ \Bigg(\frac{2 M G}{c^2 a(t)} -1 \Bigg) dr^2 +
a(t)^2 (\sin^2 \theta d\phi^2 + d \theta^2).
\label{metricb}
\end{eqnarray}
Finally a metric, in terms of two functions $a(t)$ and $b(t)$ reads: 
\begin{eqnarray}
ds^2 = - c^2 dt^2 + b^2 (t)  dr^2 + a^2 (t)  (\sin^2 \theta d\phi^2 + d \theta^2).
\label{metricab}
\end{eqnarray}
It is the metric of an homogeneous, anisotropic space with spatial
section of topology $\mathbf{R} \times \mathbf{S}^2$.  In our case,
$b(t)$ is a function of $a(t)$, $b=b(a(t))$.

\section{Dynamical equations for the black hole interior}

The action for Einstein gravity can be written in the form
\begin{eqnarray}
S = \frac{c^4}{16 \pi G} \int d^3 x dt \mathcal{N} q^{1/2} \Big[ K_{ij} K^{ij} -K^2 + ^{(3)}\mathcal{R}\Big],
\label{Action}
\end{eqnarray}
\noindent where $\mathcal{N}$ is the lapse function, $q$ is the determinant of the metric of the three-dimensional space-like hypersurfaces of ADM formalism, $K_{ij}$ is the extrinsic curvature tensor, $K$ is its trace and $^{(3)}\mathcal{R}$ is the Ricci scalar in the hypersurface.\\
\indent We can consider the Hamiltonian given by,
\begin{eqnarray}
 H=H_{0}+E_{ADM}, \nonumber
\end{eqnarray}
\noindent where $H_0$ contains the constraints of the Hamiltonian, and $ E_{ADM}$ is known as the ADM energy. This last term is an additional boundary term for asymptotically flat spacetimes. $H_0$ can be written in terms of the lapse function $\mathcal{N}$ and the shift vector $\mathcal{N}^i$ of the ADM formalism. It is given by,
\begin{eqnarray}
 H_{0} = \int d^{3}x(\mathcal{N}\mathcal{H}+\mathcal{N}^{i}\mathcal{H}_{i}), \nonumber
\end{eqnarray}
\noindent where the $\mathcal{H}$ is called the super-Hamiltonian and the $\mathcal{H}_i$ super-momentum \cite{santini5092geometrodinamica}. The ADM energy is \cite{makela1996schroedinger}:
\begin{eqnarray}
  E_{ADM} = \frac{c^{4}}{16\pi G}\oint d^{2}s_{j}(g_{ij,i}-g_{ii,j}), \nonumber
\end{eqnarray}
\noindent where $ i, j = 1,2,3 $ and $g_{ij}$ is the spacetime metric. Where we get \cite{makela1996schroedinger}:
\begin{eqnarray}
 E_{ADM}=Mc^{2}, \label{energiaBN}
\end{eqnarray}
\noindent where we will consider the equation (\ref{energiaBN}) as the energy of black hole following \cite{makela1996schroedinger}.\\
\indent Now, it becomes necessary to write the ADM energy and the Hamiltonian's constraints in terms of the variable $ a(t) $ and the canonical moment  $\Pi$. For this, we consider the Hamiltonian constraint of just for the gravitational field, given by:
\begin{eqnarray}
 \mathcal{H}=\frac{c^{4}}{16\pi G}\sqrt{q}(K_{ij}K^{ij}-K^{2}-\mathcal{R})=0, \nonumber
\end{eqnarray}
\noindent where $q$ is the determinant of the metric of a hypersurface in the ADM formalism, $ K_{ij} $ is the extrinsic curvature tensor, $ K $ is its trace, and $ \mathcal{R} $ is the Ricci scalar in the three-dimensional spacelike hypersurface .

In particular, for the metric  of the form (\ref{metricab}), the action becomes \cite{modesto2004}:
\begin{eqnarray}
S = -  \frac{c^4}{16 \pi G}\int dt  \int_0^R dr \int_0^{2 \pi} d \phi \int_0^{2 \pi} d \theta \sin \theta
 \, a^2 \, \frac{b}{c^2} \Bigg[2 \, \frac{\dot{a}^2}{a^2} + 4 \, \frac{\dot{a} \, \dot{b}}{a \, b} - \frac{2c^2}{a^2} \Bigg], \nonumber
\end{eqnarray}
\noindent which gives us,
\begin{eqnarray}
S = - \frac{c^4 R}{2 G}\int dt \Big[b \, \frac{\dot{a}^2}{c^2} + 2 \, \dot{a} \, \dot{b} \, a - b\Big], \label{Action.Mini} 
\end{eqnarray}
\noindent where $R$ is a cutoff on the space radial coordinate \cite{modesto2004}.

From (\ref{metricb}) we noticed that the two functions $a(t)$ and $b(t)$
are not independent, so
\begin{eqnarray}
b^2(t) = \frac{2 M G}{c^2 a(t)} - 1.
\end{eqnarray}
\indent Now we calculate the Hamiltonian constraint, for this, we start from (\ref{Action.Mini}). We identify the Lagrangian given by:
\begin{eqnarray}
{\cal L} =  b \, \frac{\dot{a}^2}{c^2} + 2 \, \dot{a} \, \dot{b} \, a - b.
\label{Lagrangian}
\end{eqnarray}
Using the definition of the generalized moment $p=\partial {\cal L}/\partial{\dot a} $ and the eq. (8) we can obtain the Hamiltonian constraint, in terms of the variable $a$, as

\begin{eqnarray}
{\cal H} = p_{a} \, \dot{a}+p_{b} \, \dot{b} - {\cal L} = -\frac{1}{a b} \left[ a \left(\frac{\dot a}{c}\right)^2+a-\frac{2MG}{c^2}\right].
\label{Hamiltonian}
\end{eqnarray}
It is easy to show that the solution to this equation leads us to  reproduce the inner metric of a black hole.

\section{Quantization of interior of black hole}

\indent We can rewrite the above equation (\ref{Hamiltonian}) and point out that the mass $M$ of a black hole depends on $a$ and $\dot{a}$ its time derivative, that is, rewriting the above equation to $M$, we get:
\begin{eqnarray}
 M=\frac{1}{2G}(a\dot a^{2}+c^{2}a). \nonumber
\end{eqnarray}
\indent Recalling that the energy of the black hole is $ Mc^2$, the Hamiltonian can be expressed by:
\begin{eqnarray}
 H=\frac{c^{2}}{2G}a\dot a^{2}+\frac{c^{4}}{2G}a, \nonumber
\end{eqnarray}
\noindent where the first term on the right side is associated with the kinetic term and the second term with the potential energy of the black hole. The above Hamiltonian was also found by \cite{modesto2006loop}. Expressing the Lagrangian, one has:
\begin{eqnarray}
 \mathcal{L}=\frac{c^{2}}{2G}a\dot a^{2}-\frac{c^{4}}{2G}a. \nonumber
\end{eqnarray}
\noindent Writing the canonical moment for the variable $ a $, we get:
\begin{eqnarray}
 \Pi=\frac{\partial L}{\partial\dot a}=\frac{c^{2}}{G}a\dot a. \nonumber
\end{eqnarray}
\noindent Thus the classic hamiltonian of a black hole can be expressed by:
\begin{eqnarray}\label{hamiltoniano}
 H=\frac{G}{2c^{2}}\frac{1}{a}\Pi^{2}+\frac{c^{4}}{2G}a. 
\end{eqnarray}
\indent Let us write a Schr\"odinger equation for the interior of a Schwarzschild black hole like
\begin{eqnarray}
 \hat{H}\psi=E_{ADM}\psi, \label{BN2}
\end{eqnarray}
\noindent where $\hat{H}$ represents the hamiltonian operator of the black hole, $\psi$ is the wave function and $E_{ADM}$ is the energy of the black hole. This energy can be expressed as:
\begin{eqnarray}
 E=M_{ADM}c^{2}, \nonumber
\end{eqnarray}
\noindent here $c$ is the velocity light and  $M_{ADM}$ is the black hole mass. Solving the equation (\ref{BN2}), we will find the energy's eigenvalues $E_n$ for the Hamiltonian \hyphenation{o-pe-ra-tor}operator and the eigenfunctions $\psi_n$. So the possible mass values for black holes are established.\\
\indent We can express the spectrum of the event horizon area of the Schwarzschild black hole of through the equation:
\begin{eqnarray}
 A_{s}=\frac{16\pi G^{2}}{c^{4}}M^{2}. \label{BN1.3}
\end{eqnarray}
\indent To obtain the Schr\"odinger equation for a black hole, one must transform the classical Hamiltonian into the Hamiltonian operator and the  states will be represented by  the wave functions $ \psi = \psi(a) $ for the black hole. \\
\indent In this work we will adopt the proposal given by \cite{vsteigl2009model} for the operator moment  as:
\begin{eqnarray}
 \hat{\Pi}^{2}=-\hbar^2 \frac{1}{a^{p}}\frac{\partial}{\partial a}a^{p}\frac{\partial}{\partial a}, \nonumber
\end{eqnarray}
\noindent where $ p $ represents the ordering factor of the operator.\\
\indent Thus, the Hamiltonian given by (\ref{hamiltoniano}) can be written as:
\begin{eqnarray}
 \hat{H}=\left[-\frac{\hbar^{2} G}{2c^{2}}\frac{1}{a}\left(\frac{d^{2}}{da^{2}}+\frac{p}{a}\frac{d}{da}\right)+\frac{c^{4}}{2G}a\right], \nonumber
\end{eqnarray}
\noindent by applying the above operator in the wave function $ \psi(a) $, we get:
\begin{eqnarray}
 \left[-\frac{\hbar^{2} G}{2c^{2}}\frac{1}{a}\left(\frac{d^{2}}{da^{2}}+\frac{p}{a}\frac{d}{da}\right)+\frac{c^{4}}{2G}a\right]\psi(a)=E\psi(a). \label{BN2.40}
\end{eqnarray}
\noindent This is the Schr\"odinger time independent equation for a Schwarzschild black hole. Rewriting it in terms of Planck's length $ l_{Pl}^{2}=\frac{\hbar G}{c^{3}} $ the last equation become:
\begin{eqnarray}
 \left[-\frac{{l_{{Pl}}}^{2}}{2}\frac{1}{a}\left(\frac{d^{2}}{da^{2}}+\frac{p}{a}\frac{d}{da}\right)+\frac{a}{2l_{Pl}^{2}}\right]\psi(a)=\frac{E}{c\hbar}\psi(a). \label{BN2..41}
\end{eqnarray}
\noindent Finally  the term on the right-hand side of the above equality will be rewritten in terms of the Schwarzschild radius $ R_{s}=\frac{2GM}{c^{2}} $ where $ \frac{E}{c\hbar}=\left(\frac{Mc^{2}}{c\hbar}\right)\left(\frac{2c^{3}G}{2c^{3}G}\right)=\left(\frac{2GM}{c^{2}}\right)\left(\frac{c^{3}}{2\hbar G}\right)=\frac{R_{s}}{2l_{Pl}^{2}}$ then  the equation (\ref{BN2..41}) becomes:
\begin{eqnarray}
 -l_{Pl}^{4}\left(\frac{d^{2}}{da^{2}}+\frac{p}{a}\frac{d}{da}\right)\psi(a)+(a^{2}-aR_{s})\psi(a)=0, \label{BN2.41}
\end{eqnarray}
\noindent this is the Schr\"odinger equation for black holes written in terms of the scale factor $ a(t) $, the length of Planck $ l_{Pl} $ and the Schwarzschild radius $R_{s}$.

\section{Solution to Schr\"odinger equation for black holes}

\indent In this section the Schrodinger equation for black holes given by (\ref{BN2.41}) will be solved. For this, consider initially the change of variable:
\begin{eqnarray}
 x=\frac{a}{l_{Pl}}, \nonumber
\end{eqnarray}
\noindent using these relationships, one can write the equation (\ref{BN2.41}) in terms of the $x$ variable, becoming:
\begin{eqnarray}
 -\left(\frac{d^{2}}{dx^{2}}+\frac{p}{x}\frac{d}{dx}\right)\psi(x)+\left(x^{2}-x\frac{R_{s}}{l_{Pl}}\right)\psi(x)=0, \nonumber
\end{eqnarray}
\noindent let us denote $ N = R_s / l_{Pl}$, where $ N $ will be a real number. So the above equation becomes:
\begin{eqnarray}
 -\left(\frac{d^{2}}{dx^{2}}+\frac{p}{x}\frac{d}{dx}\right)\psi(x)+\left(x^{2}-xN\right)\psi(x)=0, \label{BN2.42}
\end{eqnarray}
\noindent this equation is known as the Wheeler-DeWitt equation (WdW) for the present model, whose solution is given by:
\begin{eqnarray}
 \psi(x) &=& c_{1} e^{\frac{1}{2}(-x+N)}HeunB\left(-1+p,N,\frac{1}{4}N^{2},0,-x\right) + \nonumber \\
 &&+ c_{2} e^{\frac{1}{2}(-x+N)}HeunB\left(-1+p,N,\frac{1}{4}N^{2},0,-x\right)x^{1-p}, \label{BN2.43}
\end{eqnarray}
\noindent where $c_{1}$ and $c_{2}$ are constants and $ HeunB( \alpha,\beta,\gamma, \delta, x) $ is the solution of the biconfluent Heun equation, which is an ordinary, homogeneous, linear and second order Fuchsian \hyphenation{dif-fe-ren-tial}differential equation with four singular points \cite{ronveaux1995heun,slavjanov2000special}. The Heun equation can be written as:

\begin{eqnarray}
 \frac{d^{2}y(z)}{dz^{2}}+\left(\frac{\gamma}{z}+\frac{\delta}{z-1}+\frac{\epsilon}{z-d}\right)\frac{dy(z)}{dz}+\frac{\alpha\beta z-q}{z(z-1)(z-d)}y(z)=0, \nonumber
\end{eqnarray}
\noindent 
where $ \alpha, \beta, \gamma, \delta$, $\epsilon$ and $q$ are complex parameters and are related to the characteristic exponents of the solutions, $ d (d \neq 0,1, \infty) $ locates the fourth singular point of the equation that may or may not be complex, known as the accessory parameter. For more information on Heun's role in the context of black holes, see \cite {momeni2012quantized}. \\

\subsection{Solution close of event horizon}
Analyzing the solution given by the equation (\ref{BN2.43}), in the case where $ a \rightarrow N l_{Pl} $, that is, $ x \rightarrow N $, we set $c_{1}=0$, in order to guarantee the square integrability of the wave function and consider only the second term with $p>1$. For more details on the asymptotic behavior of the Heun function see \cite{momeni2012quantized}. Then, wave function becomes:

\begin{eqnarray}
 \psi(x) \sim c_{2} e^{\frac{1}{2}(-x+N)}HeunB\left(-1+p,N,\frac{1}{4}N^{2},0,-x\right)x^{1-p}. \label{BN2.44}
\end{eqnarray}
\indent The solution given by (\ref{BN2.44}) will be analyzed for the special case in which the third parameter of the Heun equation satisfies $ \gamma = 2 (n + 1) + \alpha $, where $ n $ is a positive integer. In this case, the $ n + 1 $ the coefficient represents in the series expansion of a polynomial of degree $n$ in $ \delta $. When $ \delta $ is the root of this polynomial and that  $ n + 1 $ coefficients and its subsequent ones cancel out, the series is truncated, resulting in a polynomial of degree $n$ for Heun \cite{ronveaux1995heun}. Applying this condition in the wave function given in (\ref {BN2.44}), we get $N= \sqrt{8(n+1)+4(1-p)}$, where $ N $ expresses the ratio between the radius of Schwarzschild and the Planck length $ N = R_s / l_{Pl} $, in this way, we have that the radius of Schwarzschild, will be $R_{s}=l_{Pl}\sqrt{8(n+1)+4(1-p)}$, where it is seen that the Schwarzschild radius is quantized and depends on the ordering factor. Here, $p$ carries the quantum information of the black hole. When you make $ p = 2 $, you retrieve the result found in \cite{makela1996schroedinger}. \\
\indent Since the area of the event horizon of a Schwarzschild black hole is determined by $A_{s}=4\pi R_{s}^{2}$, we have that,
\begin{eqnarray}
 A_{n}=32\pi l_{Pl}^{2}\left(n+\frac{3-p}{2}\right),\label{BN3.45-1}
\end{eqnarray}
\noindent 
which shows that the area of the event horizon of a Schwarzschild black hole can only assume certain discrete values of the order of the Planck length, that is, there is a minimum area defined by $ p $ and $ n $. \\
\indent Using the equation (\ref{BN1.3}) and (\ref{BN3.45-1}) it is possible to express the mass eigenvalues of a Schwarzschild black hole as $ M_{n}=M_{Pl}\sqrt{2}\sqrt{n+\frac{3-p}{2}}$, where we consider that the Planck mass $M_{Pl}=c^{2}l_{Pl}/G$. Then the energy is given by,
\begin{eqnarray}
 E_{n}=E_{Pl}\sqrt{2}\sqrt{n+\frac{3-p}{2}}, \nonumber
\end{eqnarray}
\noindent where $ E_{Pl} = \sqrt{\hbar c^5/ G}$ is the Planck energy. \\
\indent As shown by Hawking \cite{hawking1975particle}, black holes can emit radiation, and by emitting this radiation they lose mass. Thus, Hawking's radiation will be calculated for this model. Consider that the area given by the equation (\ref{BN3.45-1}) changes from $ A_n$ to $dA_n$ and the mass changes from $ M_{n} $ to $dM_{n} $, such that $dA_{n} = 32\pi l_{Pl}^{2}dn$, but we have that $dn = 1$, so the equation becomes $dA_{n} = 32\pi l_{Pl}^{2}$. Rewriting this equation in terms of the infinitesimal element $dM_{n}$ of mass, we obtain $
 dM_{n}=\frac{M_{Pl}}{\sqrt{2}}\left(n+\frac{3-p}{2}\right)^{-\frac{1}{2}}$, thus the energy emitted by a black hole in a state transition is given by:

\begin{eqnarray}
 dE_{n}=\frac{E_{Pl}}{\sqrt{2}}\left(n+\frac{3-p}{2}\right)^{-\frac{1}{2}}, 
\end{eqnarray}
\noindent or $dE_{n}=E_{Pl}^{2}/E_{n}$, where such a quantum energy of Hawking radiance depends on the ordering factor. The frequency emitted by this radiation is given by:

\begin{eqnarray}
 \nu_{n}=\frac{E_{Pl}^2}{h\sqrt{2}}\left(n+\frac{3-p}{2}\right)^{-\frac{1}{2}}, \label{frequencia}
\end{eqnarray}
\noindent it can be seen that the frequency emitted by equation (\ref{frequencia}) is quantized, since it depends on $n$ which is an integer and $p$ which represents the ordering factor of the operator. A brief discussion on frequency differences can be seen in \cite{makela1996schroedinger}.

\subsection{Solution close of singularity $a\rightarrow0$}

In this section we will analyze regions close to the interior singularity of the Schwarzschild black hole in which the scale factor is very small, $ a \rightarrow 0 $, which implies that $ x \rightarrow 0 $. Thus, one can write equation (\ref{BN2.42}) as:

\begin{eqnarray}
 \left(\frac{d^{2}}{dx^{2}}+\frac{p}{x}\frac{d}{dx}\right)\psi(x)=0, \label{BN3.850}
\end{eqnarray}
\noindent whose solution is given by:
\begin{eqnarray}
 \psi(a)=\varphi_{0}+i\varphi_{1}\left(\frac{a}{l_{Pl}}\right)^{1-p}, \label{BN3.50}
\end{eqnarray}
\noindent where $ p \neq 1 $, this restriction will be justified next, when the Bohmian trajectories are found.\\
\indent Using the DeBroglie-Bohm interpretation, one can write the phase of the wave function (\ref{BN3.50}) as,
\begin{eqnarray}
 S(a)=\tan^{-1}\left[\frac{\varphi_{1}}{\varphi_{0}}\left(\frac{a}{l_{Pl}}\right)^{1-p}\right], \nonumber
\end{eqnarray}
and the real part of the wave function (\ref{BN3.50}) will be:
\begin{eqnarray}
 R(a)=\sqrt{\varphi_{0}^{2}+\varphi_{1}^2 \left(\frac{a}{l_{Pl}}\right)^{2(1-p)}}. \nonumber
\end{eqnarray}
The general equation that gives the Bohm's trajectory, may be written for the scale factor, in the present model, as,
\begin{eqnarray}
\dot{a}=-\frac{1}{a}\frac{\partial S}{\partial a}. \nonumber
\end{eqnarray}
Solving that equation, the bohmian trajectories are given by,
\begin{eqnarray}
a(t) = \left\{ \begin{array}{ll}
\left[\left(\frac{\varphi_{1}}{\varphi_{0}}\right)l_{Pl}^{-1+p}(1-p)(|1-p|-3)(t+t_{0})\right]^{\frac{1}{3-|1-p|}}, & \textrm{for $|1-p|\neq0,3$},\\
e^{-\left(\frac{\varphi_{1}}{\varphi_{0}}\right)l_{Pl}^{-1+p}(1-p)(t-t_{0})}, & \textrm{for $|1-p|=3$}, \label{BN3.100}
 \end{array}\right.
\end{eqnarray}
\noindent when $ | 1-p | \neq 0,3  $, the Bohm's trajectory $ a (t) $ is finite and regular, \hyphenation{con-si-de-ring}considering very small values of $ a $, which results in the removal of the singularity. For the case where $ | 1-p | = 3 $, the exponential behavior was obtained. 

Through the quantum potential proposed by De Broglie-Bohm theory, we can understand why the scale factor does not go to zero when $t\rightarrow0$. To determine the quantum potential we consider the following expression \cite{oliveira2018broglie}:

\begin{eqnarray}
Q(a)=-\left(\frac{1}{R(a)}\frac{\partial^{2}R(a)}{\partial a}-\frac{p}{a}\frac{1}{R(a)}\frac{\partial R(a)}{\partial a}\right), \label{npq}
\end{eqnarray}
the second term of equation (\ref{npq}) is related to the ambiguity of the quantum operator ordering factor choice \cite{colistete1998singularities,he2014spontaneous}. Using the equation for the real part of the wave function, we can express the quantum potential as:

\begin{eqnarray}
Q(a)&=& -\frac{(-1+p)\varphi_{1}^2}{\left(\varphi_{0}^2l_{Pl}^2\left(\frac{a}{l_{Pl}}\right)^{2p}+\varphi_{1}^2a^2\right)^3} \nonumber \\
&&\times \bigg[3\varphi_{0}^{4}p\left(\frac{a}{l_{Pl}}\right)^{4p}l_{Pl}^{4}+5\varphi_{0}^{2}\varphi_{1}^{2}p\left(\frac{a}{l_{Pl}}\right)^{2p}a^{2}l_{Pl}^{2}-\varphi_{0}^{4}\left(\frac{a}{l_{Pl}}\right)^{4p}l_{Pl}^4  \nonumber \\
&&-\varphi_{0}^2\varphi_{1}^2\left(\frac{a}{l_{Pl}}\right)^{2p}a^2l_{Pl}^2+2\varphi_{1}^4a^4p \;\;\;\bigg]. \label{npq1}
\end{eqnarray}

\begin{figure}[!htb]
    \centering
    \includegraphics[scale=0.5]{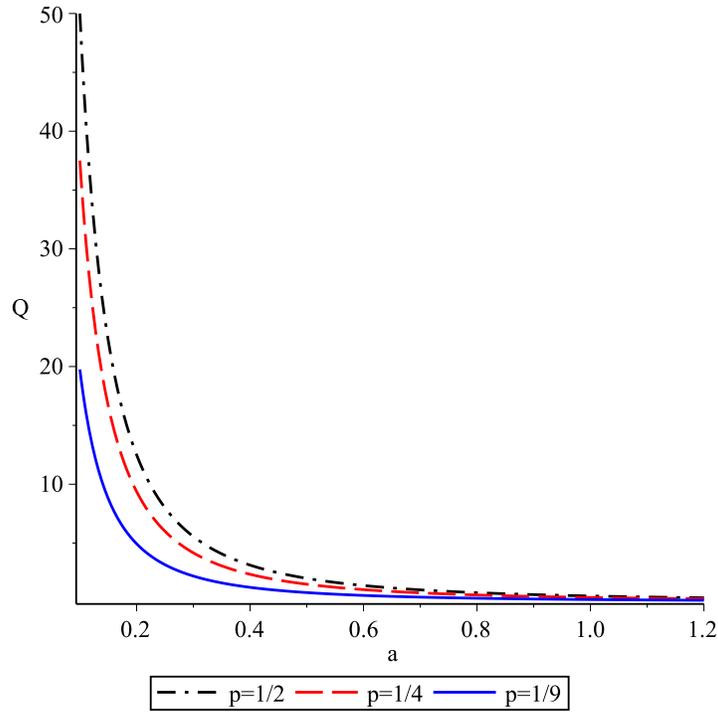}
    \caption{Quantum potential $Q$ providing the De Broglie-Bohm trajectory for the values of $p=1/2;1/4;1/9$.}
    \label{figpotquan}
\end{figure}

\indent After a detailed analysis of the expression given by (\ref{npq1}), we find that for some values of $ p $ the quantum potential is positive and finite, as $ a $ increases, the quantum potential decreases and goes asymptotically to zero. when $a\rightarrow0$. Some of these cases can be seen in the figure (\ref{figpotquan}).

\section{Conclusions}
\indent 
In this work we have studied the interior of a Schwarzschild black hole using the metric (5) in terms of the scale factor $ a $, the Planck length $ l_{Pl}$ and the Schwarzschild radius $R_{s}$. We obtained equation (\ref{BN2.42}). The wave function found is written in terms of the Heun function, which is a solution of the Heun biconfluent equation. The behavior of this wave function was analyzed close of the horizon, in the case  it is possible to rewrite it in the form given by the equation (\ref{BN2.44}). We investigated the special case in which the third parameter of the Heun equation satisfies the condition $ \gamma = 2 (n + 1) + \alpha $ where $ n $ is a positive integer. It was possible to quantify results for the case where the Schwarzschild radius coincides with the event horizon and consequently find the spectrum of area, mass, energy and the frequency emitted by Hawking radiation. It was possible to quantify the event horizon of the Schwarzschild, black hole and it should be noted that all quantised quantities depend on the ordering factor of the operators, which had not been considered yet, in the literature.\\
\indent Then the singularity inside the black hole was analyzed. For this, it was considered that $a\rightarrow 0$ in equation (\ref{BN2.41}), which was written as it appears in (\ref{BN3.850}). The wave function is given by equation (\ref{BN3.50}), which, when interpreted from DeBroglie-Bohm's point of view, resulted in solution (\ref{BN3.100}), which describes how space-time evolves over time. This solution depends on the ordering factor of the operator $p$, which will have restrictions for some values, since they do not lead to solutions that have physical meaning. So, for the case where $| 1-p |\neq 0,3 $, the Bohm's trajectory is finite and regular, that is, the singularity is removed. For the case where $| 1-p |=3$, the Bohm's trajectory assumes an exponential behavior, never going to zero, avoiding the singularity and allowing space-time to be extended beyond the classical singularity.\\
\indent In summary, we have presented a mathematical proof that the Schwarzschild horizon is quantized an the interior singularity \hyphenation{di-sap-pears}disappears. This same procedure  can be extended to other black holes such, Reisner-Nordström and Kerr. It will be studied in our future work. It is interesting to observe that metric will go beyond the classical singularity, like recently was observed \cite{bianchi2018white}.

%\newpage
%\bibliographystyle{unsrt}
%\bibliography{refs}
\section{References}
\noindent 
\bibliographystyle{unsrt}
\bibliography{BuracoNegroLKN}



%% References


%\bibliography{sample.bib}






% \end{thebibliography}

\end{document}